# PARAM: A Microprocessor Hardened for Power Side-Channel Attack Resistance

Muhammad Arsath K F, Vinod Ganesan, Rahul Bodduna, and Chester Rebeiro
Department of Computer Science and Engineering
Indian Institute of Technology Madras, India
{muhammadarsath,vinodg,rahulb,chester}@cse.iitm.ac.in

***Abstract*—The power consumption of a microprocessor is a huge channel for information leakage. While the most popular exploitation of this channel is to recover cryptographic keys from embedded devices, other applications such as mobile app fingerprinting, reverse engineering of firmware, and password recovery are growing threats. Countermeasures proposed so far are tuned to specific applications, such as crypto-implementations. They are not scalable to the large number and variety of applications that typically run on a general purpose microprocessor.**

**In this paper, we investigate the design of a microprocessor, called PARAM with increased resistance to power based side-channel attacks. To design PARAM, we start with identifying the most leaking modules in an open-source RISC V processor. We evaluate the leakage in these modules and then add suitable countermeasures. The countermeasures depend on the cause of leakage in each module and can vary from simple modifications of the HDL code ensuring secure translation by the EDA tools, to obfuscating data and address lines thus breaking correlation with the processor's power consumption. The resultant processor is instantiated on the SASEBO-GIII FPGA board and found to resist Differential Power Analysis even after one million power traces. Compared to contemporary countermeasures for power side-channel attacks, overheads in area and frequency are minimal.**

## I. Introduction

The Internet of Things (IoT) is becoming ubiquitous and revolutionizing society at large. It is estimated that there will be around 20 billion connected IoT devices by 2020 [1]. However, embedded processors, which form a core component of an IoT device, are highly vulnerable to power side-channel attacks [2]. These attacks glean sensitive information leaked by exploiting the correlation between the sensitive data and the processor's power consumption. The attacks have been extensively used to break crypto algorithms and recover secret keys [3]. More recently, they have been applied to reverse engineer firmware [4], recover passwords [5] and fingerprint mobile apps [6]. Traditionally, power attacks were countered by the use of application level mitigations such as masking the device's power consumption [7] or hardware level leakage hiding mechanisms [8]. However, these strategies have considerable overheads and increase the area and power requirements by orders of magnitude. Thus their use is restricted to a few specialized components of the system, such as crypto modules, leaving the remaining components still vulnerable.

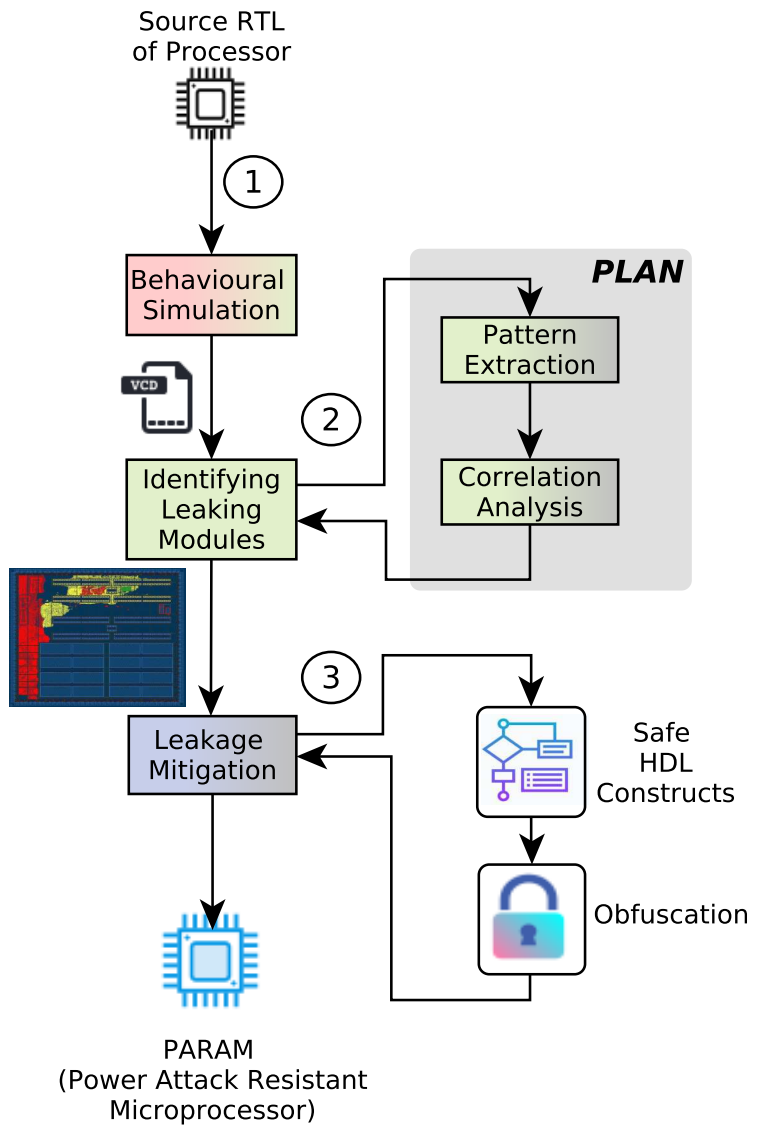

Fig. 1: To design a power-attack resistant microprocessor, we start with the RTL of a processor, use behavioral simulation to identify hotspots of leakage, and then apply suitable countermeasures to these locations.

In this paper, we take a complementary approach, where we fundamentally investigate the design of an embedded microprocessor, called PARAM (an acronym for Power Attack ResistAnt Microprocessor), with significantly less leakage through power channels. To achieve this, we (1) evaluate the side-channel vulnerability of an open-source processor design, (2) identify modules that leak the most, and (3) apply appropriate countermeasures to the leaking modules. The resultant is a processor with significantly increased side-channel resistance. This approach universally safeguards all the applications executing on the processor. Also, as this is a ground-up design process, the processor has negligible performance degradation, as well as low power and area overheads as opposed to standard power attack protection mechanisms such as masking [7], [9] and hiding [10], [11].

To facilitate such a processor design, we first develop a framework called PLAN: Power attack Leakage ANalyzer, which, at design time can provide a quick estimate of the side-channel leakage of a microprocessor. The PLAN framework takes the source RTL of a processor architecture and benchmark applications as input. It provides a report illustrating the amount of information leakage from each processor module through the processor's power consumption.

We apply PLAN to a reference open-sourced RISC-V embedded processor architecture, Shakti-C [12], and identify the modules with high leakage. These include the register bank, cache memory, control status registers, execute stage units and pipeline buffers. We also find that the fixed mapping from address to cache sets is a major source of leakage. Further, we found that the EDA tools used in source RTL translation introduced certain translations, which were functionally correct, but increased the side-channel leakage of the processor. Due to this, Floating Point and Branch Prediction Units leaked information about an AES secret key. These results are surprising because the evaluated AES implementation did not have any floating point operations nor any branches that were dependent on the secret key.

To strengthen the processor against power attacks, we evaluate the cause of the leakage and fix it module by module. To achieve this, we first ensure that EDA translations are not just functionally correct but also do not increase side-channel vulnerability. Further, we ensure that any data present in the processor's data path is obfuscated. Thus, data present in the cache memory, line buffers, general purpose registers, and pipeline buffers is in an obfuscated form. The obfuscation breaks the correlation between the data and the power consumption, thus reducing leakage. De-obfuscation only happens in the execute stage of the processor, when an operation needs to be performed on the data. The result of the operation is obfuscated again before writing back to the registers. Thus, in the enhanced processor, PARAM, this mapping is also obfuscated. Obfuscation is achieved by a 4 round Feistel function with a key that is generated and hidden inside the processor. The key is periodically changed to ensure that the scheme is not easily broken.

To demonstrate the efficacy of our approach, we compare Differential Power Analysis (DPA) [2] on the original Shakti-C processor and PARAM. The target application is a software implementation of AES-128. We found that on average, DPA was able to recover the AES-128 secret key within 62K power traces on the baseline Shakti-C. On the other hand, DPA failed to reveal any bytes of the secret key even after one million power measurements on PARAM. The increase in area and energy consumed by the processor is small compared to contemporary side-channel countermeasures.

In summary, our key contributions are as follows:

- We propose a framework called PLAN to provide a quick indication of leaking modules in microprocessors. We use PLAN to identify the leaking modules in an open-source RISC V processor, Shakti-C [12] and show that cache memories, register files, ALU, and pipeline buffers are a major cause of leakage.
- We experimentally show that performance-centric translations performed by EDA tools can increase the processor's side-channel leakage as a side-effect.
- We provide an extensive evaluation of side-channel leakage through cache memories. Our results show that cache memories leak considerably more about the memory address than the data stored in the cache. Due to this, the memory address used in a load/store instruction is more vulnerable to side-channel leakage than the data fetched/stored.
- We demonstrate that a combination of HDL programming and obfuscation techniques can be used to design processors with significantly reduced power side-channel leakage. The performance, area, and energy overheads to achieve this side-channel security are minimal.
- We evaluate PARAM on the SASEBO-GIII platform and demonstrate the resilience to Differential Power Attacks.

The paper is organized as follows. Section II explains the background on Differential Power Analysis and a side-channel vulnerability metric used by PLAN. In Section III, details of the PLAN framework is explained and the results of leakage analysis on the reference processor are briefly discussed. Section IV presents the steps involved in designing PARAM, power side-channel attack resistant microprocessor. Section V presents the leakage results and DPA results on PARAM. In Section VI, related work on various countermeasures are discussed and the limitations of PARAM are enumerated in Section VII. The final section concludes the paper.

## II. Background

In this section, we provide a brief introduction to Differential Power Analysis (DPA) and a metric called Side Channel Vulnerability Factor (SVF) [13], which is used in PLAN.

### A. Differential Power Analysis

In a Differential Power Analysis (DPA) [2], an attacker extracts secret information from a device by correlating its power consumption patterns with a hypothetical model of the device's power consumption [14]. The most commonly used power consumption models are Hamming distance and Hamming weight. While the former computes the number of bit toggles between two consecutive values held by a register, the latter captures the number of bits that are set in a register. Typical countermeasures for DPA try to break this correlation by randomizing or hiding the power consumption patterns. However, all countermeasures proposed so far are mostly targeted toward crypto-algorithms.

### B. Side-channel Vulnerability Factor

Side-channel Vulnerability Factor (SVF) [13] is a metric used to quantify the side-channel leakage of a device. It measures the correlation between a sensitive application's (called victim) execution pattern and the attacker's side-channel observations. A high correlation indicates a strong side-channel leakage. Computing the SVF involves two phases: an online phase and a post-processing phase.

In the online phase, the application is triggered with $N$ different inputs. For each input, an *Oracle trace*, is built to contain the ground truth that an adversary ideally wants to read from the device. The Oracle trace can be denoted as

$$\mathbb{O} = (o_1, o_2, o_3, \ldots, o_i, \ldots . o_N) \ , \tag{1}$$

where $o_i$ $(1 \leq i \leq N)$ is the ground truth for the $i$-th input. For example, for an AES victim, the input or key is varied and execution is triggered. The Oracle Trace may contain values such as $S\,(p \oplus k)$, where $p$ is a byte of the plaintext, $k$ the corresponding secret key byte, and $S$ the AES S-box operationDuring each execution, the device's power consumption patterns are collected and stored in a *Side-Channel trace* as follows:

$$\mathbb{S} = (s_1, s_2, s_3, \ldots, s_j, \ldots s_N) \ , \tag{2}$$

where $s_j$ is the power trace collected for the $j$-th input.

At the end of the online phase, we have two series of data. The post-processing phase identifies patterns in each of the traces and then computes the correlation between them. Patterns are identified by computing pairwise distances between data in each trace using a distance metric. This leads to an Oracle Distance Vector $\mathbb{D}_O$ having $\binom{N}{2}$ values defined as follows:

$$\mathbb{D}_O = \big(distance(o_i, o_j), \quad \forall o_i, o_j \in \mathbb{O} \text{ and } i > j\big) \ .$$

Similarly, a Side-channel Distance Vector $\mathbb{D}_S$ having $\binom{N}{2}$ values is constructed from $\mathbb{S}$ using a distance metric. Pearson's correlation is then computed between these two vectors and the correlation score is interpreted as the SVF. A high SVF indicates that the Oracle leaks significantly through the side-channel. The distance metric used depends on the input data. In this paper, we use the Hamming distance metric to find distances between binary data. The Hamming distance is defined as follows:

$$HD(x_i, x_j) = \sum_{b=0}^{n-1} h_b \in \mathbb{N} \ , \tag{3}$$

where $x_i$ and $x_j$ are $n$-bit binary strings and $h_b$ is the $b$-th bit of $(x_i \oplus x_j)$.

## III. Identifying Power Side-Channel Leakage in Microprocessors

To improve the side-channel resiliency of a processor, we start with the processor source code (RTL) and identify components that leak side-channel information. We then incorporate countermeasures for the leaking modules. In this section, we present a framework called Power Attack Leakage ANalyzer PLAN, which works on the RTL of the processor to identify modules with high leakage.

The processor is represented as a netlist of functional modules. For example the RISC V processor, Shakti-C [12] has 430 modules such as the main pipeline, FPU, BPU, ALU, data cache, instruction cache, etc. Each module consists of sub-modules and signals comprising of a set of wires and registers. PLAN evaluates every module in the processor independently. For a given module, it estimates leakage from the signals associated with that module. Leakage due to sub-modules is estimated separately.

Formally, we denote the processor netlist, $P$, as a set of all modules, i.e.

$$P = \{M_1, M_2, \ldots, M_i, \ldots, M_m\} \ ,$$

where $M_i$ $(1 \leq i \leq m)$ is the $i$-th module. Each module comprises of signals (wires and registers), which we denote by the set

$$M_i = \{S_1^{(i)}, S_2^{(i)}, \ldots, S_j^{(i)}, \ldots, S_n^{(i)}\} \ ,$$

where $S_j^{(i)}$ $(1 \leq j \leq n)$ is a signal defined in the module $M_i$ and can be of one or more bits.

PLAN is designed based on the following assumptions:

1) Power consumed by a $k$-bit signal, $S$, is proportional to its Hamming weight. We define the power model by the function
$$\mathsf{P}_M(S) = \sum_{b=0}^{k-1} S_b \ ,$$
where $S_b$ is the $b$-th bit in $S$.
2) Let $\mathsf{Y}(M_i)$ be a function that concatenates all signals in $M_i$. We define this function as follows:
$$\mathsf{Y}(M_i) = (S_1^{(i)} || S_2^{(i)} || \ldots || S_n^{(i)}) \ ,$$
where $||$ represents concatenation operation. Thus, if each signal in $M_i$ is 32-bits, $\mathsf{Y}(M_i)$ will be 32 $\times n$ bits. We estimate the power consumed by module $M_i$ as an aggregation of the power consumed by all signals in module $M_i$. Thus, the data held in $\mathsf{Y}(M_i)$ directly relates to the power consumed by the module $M_i$.
3) SVF between $\mathsf{Y}(M_i)$ and the secret data provides an estimate of the side-channel leakage for the module $M_i$.

PLAN works by executing benchmark programs multiple times with different inputs on a pre-synthesized netlist. Before starting the execution, the user identifies one or more interesting operations in each benchmark program. For example, some of the interesting operations in the first round of an AES benchmark program are $p \oplus k$, $S(\,p \oplus k\,)$, $2 \cdot S(\,p \oplus k\,)$ and so on. The contents of these locations are used as Oracle trace. To quantify side-channel leakage from a module $M_i$, the side-channel trace is formed by capturing data present in $\mathsf{Y}(M_i)$. For each execution of the benchmark, $\mathsf{Y}(M_i)$ is a time-series vector whose elements correspond to data in each clock cycle. SVF between the Oracle trace and side-channel trace is computed by first performing a pattern extraction and then a correlation analysis as discussed below.

**Pattern Extraction.** Assume that each benchmark is executed $N$ times and each run is for $d$ clock cycles. The data present in $\mathsf{Y}(M_i)$ in the $j$-th run of the benchmark forms a vector of the form $(y_{(j,1)}, y_{(j,2)}, y_{(j,3)}, \ldots, y_{(j,d)})$. This vector forms an element in the side-channel trace. Thus in Equation 2,

$$s_j = (y_{(j,1)}, y_{(j,2)}, y_{(j,3)}, \ldots, y_{(j,k)}, \ldots, y_{(j,d)}) \ ,$$

where $y_{(j,k)}$ represents the data in $\mathsf{Y}(M_i)$ in the $k$-th clock cycle. We thus have a matrix of dimension $N \times d$ which forms the Side-channel trace (see Figure 2). The Oracle trace are scalars obtained from the output of an interesting point. Thus in Equation 1,

$$o_j = \langle \text{output of interesting point, such as } p \oplus k \text{ or } S(p \oplus k) \rangle$$

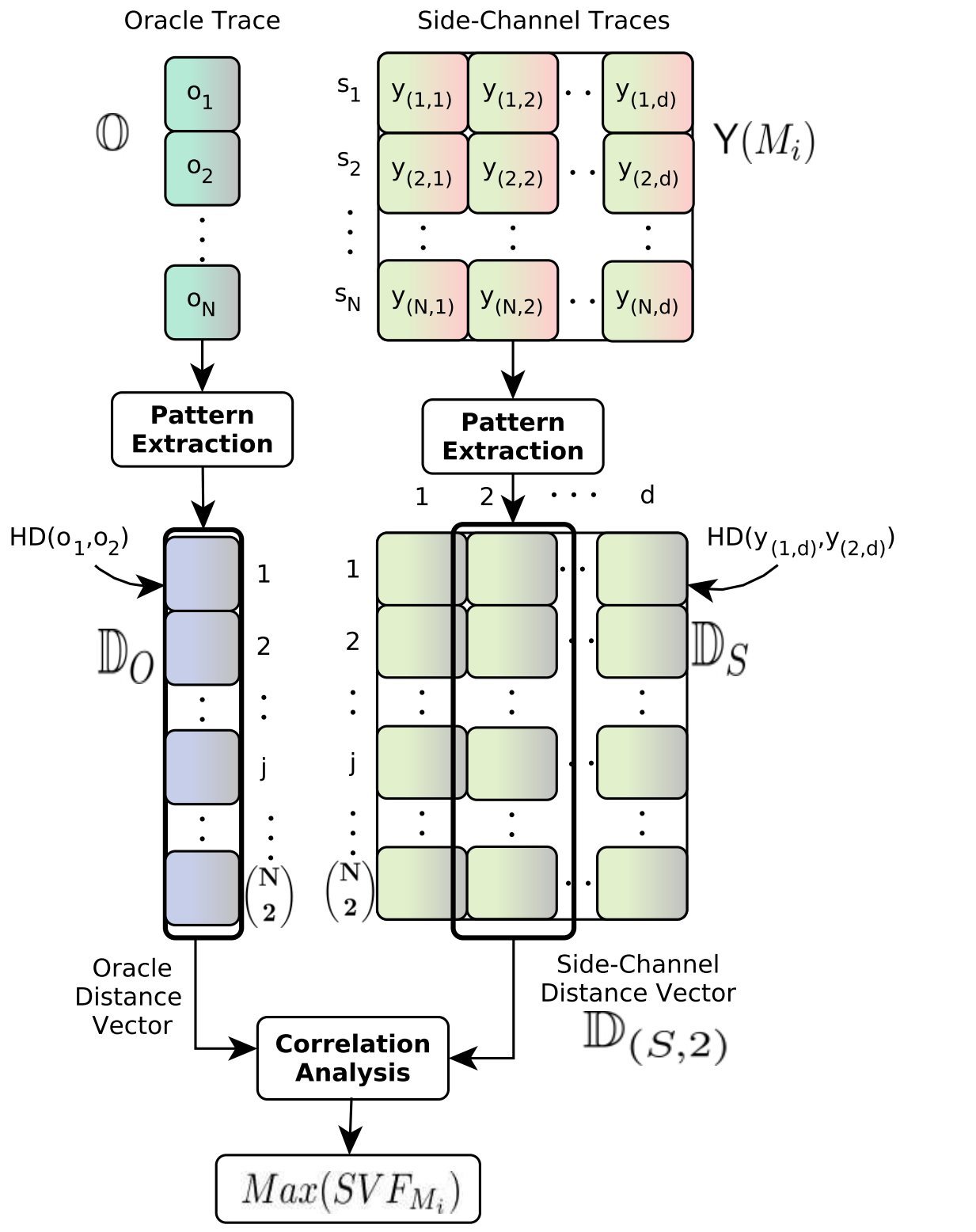

Fig. 2: Overview of SVF computation for a processor module $M_i$ using PLAN.

The next step is to identify patterns in each trace using pairwise distances to form the Oracle Distance Vector $\mathbb{D}_O$ and Side-Channel Distance Vector $\mathbb{D}_S$ (refer Section II-B). We evaluated several distance metrics, such as Euclidean distance, Cosine similarity, Hamming weight and found Hamming distance as the most suited. Hamming distance (Equation 3) captures the distance between two $n$-bit binary vectors and it performs well in the presence of noise in the signals.

**Correlation Analysis.** Each clock cycle would result in a different Side-Channel Distance Vector, which we denote $\mathbb{D}_{(S,c)}$, where $1 \leq c \leq d$. Pearson's correlation coefficient ($\rho$) is computed between $\mathbb{D}_O$ and each $\mathbb{D}_{(S,c)}$, and the absolute value is taken to provide the SVF. We thus have $d$ different SVF values, of which the maximum is considered as leakage of the module $M_i$. This is formally stated as follows:

$$SVF_{M_i} = Max(|\rho(\mathbb{D}_O, \mathbb{D}_{S,c})|) \quad 1 \leq c \leq d \ . \qquad (4)$$

### A. *Leakage Analysis on* Shakti-C

As a case study, we apply PLAN to an open-source RISC-V processor, Shakti-C [1]. Shakti-C is a 64-bit, 5-stage in-order processor with a 16KB instruction cache and a 16KB data cache. It has a bi-modal Branch Prediction Unit (BPU) with a coupled Branch Target Buffer (BTB) and an IEEE 754 compliant Floating Point Unit (FPU).

[1]https://bitbucket.org/casl/c-class/

The source code is built of 430 modules arranged hierarchically. The modules can be considered to be part of either the data path or the control path. In this work, we evaluate leakage due to the data path comprising of the data cache, register file, load-store units, and all modules in the execute and writeback stage of the processor pipeline.

We found that an implementation of AES-128 uses all of these processor components. The AES-128 implementation considered uses a 256-byte lookup table to implement the `SubBytes` operation. Separate functions are present to implement `ShiftRows`, `MixColumns`, and `AddRoundKey` operations. These operations comprise of data memory accesses and integer ALU operations. Side-channel leakage due to control path components like instruction cache, fetch and decode stages in the pipeline, and branch prediction units are not easily protected at the architecture level. Section VII provides further details.

In the first stage of PLAN, behavioral simulation of Shakti-C with the benchmark programs is performed. Value Change Dump (VCD) samples are collected with multiple random inputs given to the benchmark program. Each VCD sample captures the signals from all 430 functional modules. The PLAN framework evaluates SVF for these VCD samples to pinpoint leaking modules in Shakti-C. For the Oracle trace, all intermediate operations in the first round of the AES-128 were considered. $SVF_{M_i}$ (Equation 4) is computed for all intermediate operations, and for all the 430 modules. Table I provides the maximum $SVF_{M_i}$ observed for `SubBytes` operation. As is seen in the table, memory hierarchy comprising of the data cache, data line, and hit buffers, leak the most. Considerable leakage is also observed in the Registers and Functional Units. We found the Floating Point Unit (FPU), Branch Prediction Unit (BPU), and the Multiply-divide module in the ALU to leak considerably. This result is surprising because, the AES-128 benchmark program used does not have any floating point operations, branches, nor performs any multiplication or division. Later in this section, we investigate the cause of this leakage.

Figure 3 shows a pictorial view of the Shakti-C floor plan obtained from Synopsys IC Compiler (version M-2016.12-SP5-4). Each color denotes the magnitude of leakage, with modules in Red having the most leakage. In these modules, the secret can be completely observed through the side-channel. The modules in Blue, on the other hand, leak the least. As can be seen from Figure 3, leakage is due to a small portion of the chip. It is sufficient to insert countermeasures to protect only these regions. In the remainder of this section, we evaluate the most leaking modules in Shakti-C.

**Leakage in Memory Hierarchy.** The memory hierarchy and storage structures (cache memories, register-files, etc.) of a processor contribute considerably to data-movement. When a cache miss occurs, 64 bytes of data are loaded from the off-chip RAM to a Line Buffer (LB). The critical word is also stored in the Physical Register File (PRF) and MEM-WB buffer (refer Figure 4). The PRF is a register file used for operand forwarding while MEM-WB is a pipeline buffer between the memory stage and the writeback stage in the processor. In the next clock cycle, the data is forwarded to the data cache where it is stored in one of the 128 cache lines present. The choice of the cache set depends on the address used in the load or store instruction. When a memory operation results in a cache hit, the LB, Hit Buffer (HB) and data cache are queried in parallel. If the required data is present in any of these structures, it is returned to the MEM-WB buffer. In the following cycle, the required data is copied into the HB, if it is not already present.

TABLE I: Module wise SVF for AES-128 benchmark on Shakti-C. Leaking signals specifies the number of signals in a module that leaks, while leak count specifies the number of points in the $\mathbb{D}_S$ that have SVF more than 0.3.

| | Module | Leak Count | SVF | #Leaking Signals |
|---|---|---|---|---|
| Memory Hierarchy | Data Cache<br>Hit Buffer<br>Line Buffer | 165<br>35<br>2 | 0.99<br>0.35<br>1.0 | 85 |
| Registers | Register File<br>PRF<br>CSR | 3<br>25<br>13 | 1.0<br>1.0<br>0.64 | 24 |
| Functional Units | ALU<br>FPU<br>Mul-Div Unit | 38<br>3<br>2 | 0.99<br>1.0<br>1.0 | 21 |
| Interstage Buffers | IF-ID<br>ID-EXE<br>EXE-MEM<br>MEM-WB | 0<br>11<br>6<br>4 | 0<br>1.0<br>1.0<br>1.0 | 5 |
| Instruction Fetch | BPU | 2 | 0.95 | 1 |
| Instruction Memory | Instruction Cache<br>Line Buffer | 0<br>0 | 0.0<br>0.0 | 0 |
| TLB | Instruction TLB<br>Data TLB | 0<br>0 | 0.0<br>0.10 | 0 |

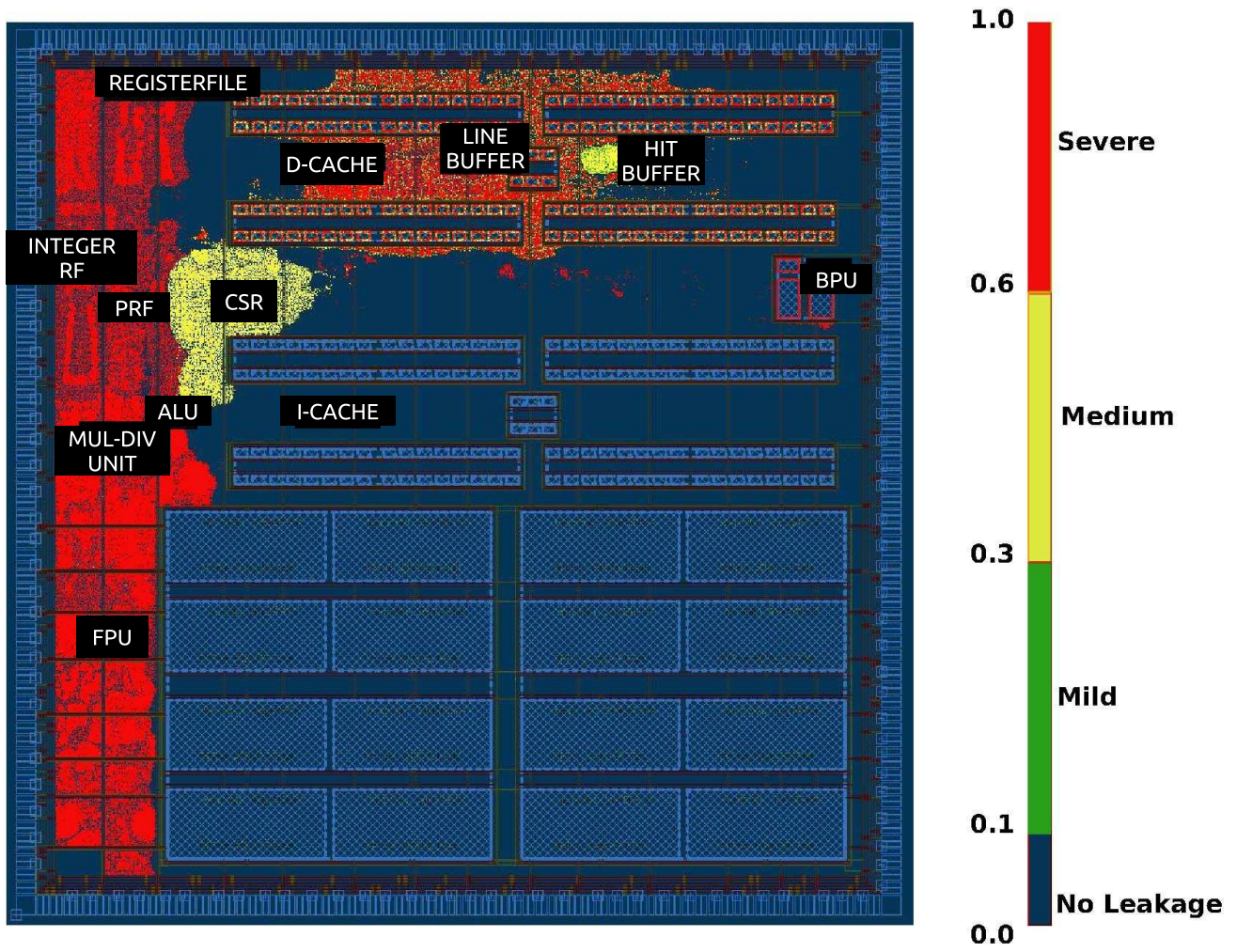

Fig. 3: Information leakage plot shows the floor plan for Shakti-C illustrating the modules with their side-channel leakage.

Every load or store operation containing data correlated with the Oracle results in a high SVF in the data cache, LB, HB, and MEM-WB buffer (Table I). For example, in the AES-128 benchmark implementation, several operations can cause leakage in these components. The `SubBytes` operation $S(p \oplus k)$ for instance, ex-ors a plaintext byte ($p$) to a key byte ($k$) and uses the result as an index into a lookup table

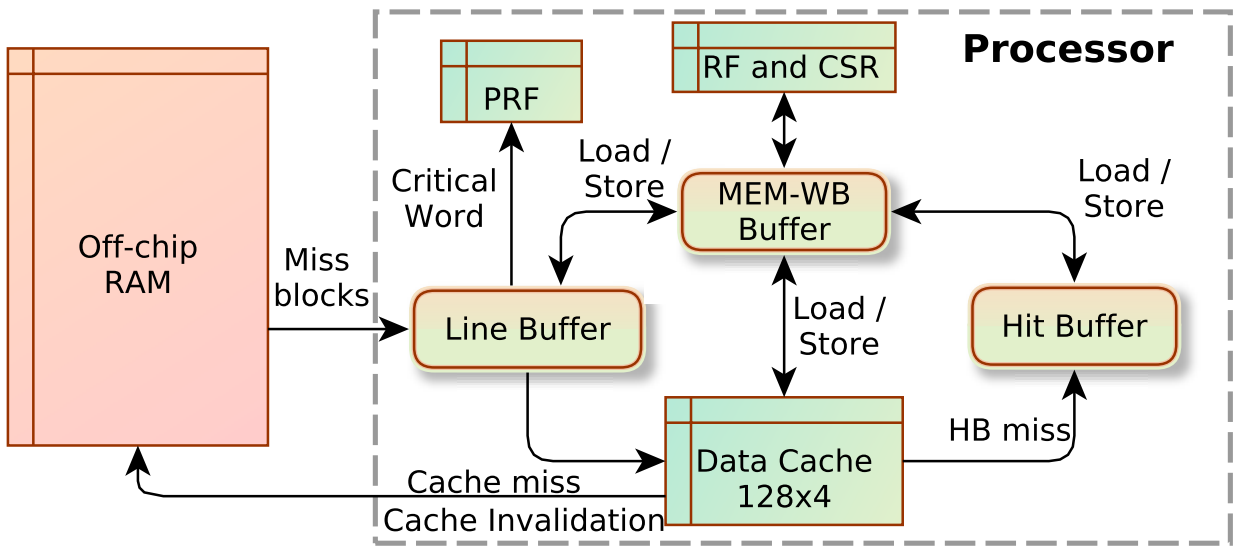

Fig. 4: Data paths of Shakti-C interconnects the storage structures.

```
                    SubBytes
1 addi   a1,gp,-1264  # a1=&p
2 lbu    a2,0(a1)     # Load value at a1 into a2
3 addi   a3,gp,-2024  # a3=&k
4 lbu    a4,0(a3)     # Load value at a3 into a4
5 xor    a5,a2,a4     # Compute p ⊕ k
6 addi   a6,gp,1332   # a6=&S
7 add    a7,a5,a6     # a7=a5+a6
8 lbu    a8,0(a7)     # Load S[p ⊕ k] into a8
```

Fig. 5: AES SubBytes operation having data memory access and integer ALU instructions .

present in the RAM. RISC V instructions to perform this is shown in Figure 5. To evaluate this code snippet, we build multiple Oracle traces: $k$ (contents of register `a4` in Line 4), index for the lookup table $p \oplus k$ (register `a7` in Line 8), and result of SBox lookup, $S(p \oplus k)$ (register `a8` in Line 8). All components in the memory hierarchy leak considerably during these load operations independent of whether there is a cache hit or a miss. The leakage is due to either the address or the data lines in the memory hierarchy. The next part of the section evaluates the leakage due to the address and data lines in the cache memory.

*Evaluating the Leakage due to Address and Data in the Cache.* An address sent to the data cache affects the tag memory and determines the cache set. On the other hand, the data only affects a specific cache line. Intuitively, since address lines influence a larger portion of the cache memory, it should leak more. We validate this with two experiments. In the first experiment, we fill all cache lines with the same data, perform multiple memory accesses to different addresses that hit in the cache and measure the differential power consumption estimates using the Hamming distance model and apply the Welch t-test [15]. This provides leakage exclusively due to address lines in the cache memory. In the second experiment, we measure the differential power consumption by accessing the same address multiple times by repeatedly modifying the data. This provides an estimate of the leakage due to data lines in the cache memory.

Figure 6a shows the t-scores between pairs of cache set (the result of the first experiment), while Figure 6c shows the t-scores between pairs of data values (the result of the second experiment). The distribution in all cases is normal, with different mean and variance (see Figure 6b and 6d). The leakage due to the address lines is considerably higher. Thus, to reduce side-channel leakage from the cache memory, it is required to obfuscate the address as well as the data lines.

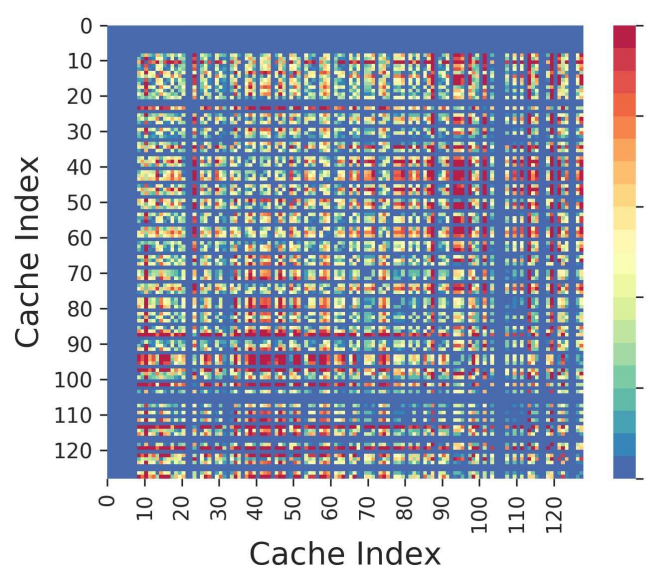

(a) Pairwise differential power consumption of cache sets.

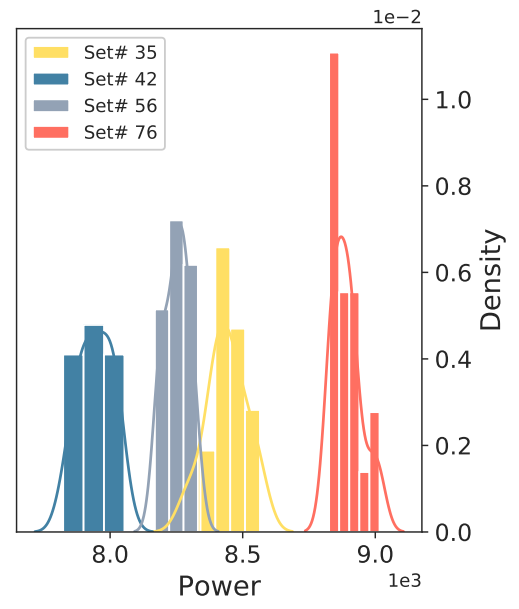

(b) Power consumption distribution for 4 arbitrarily chosen cache sets.

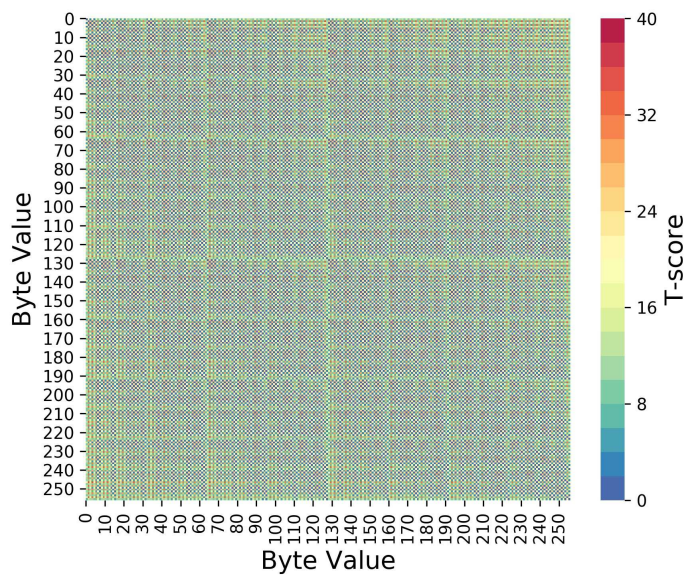

(c) Pairwise differential power consumption for data stored in a cache set.

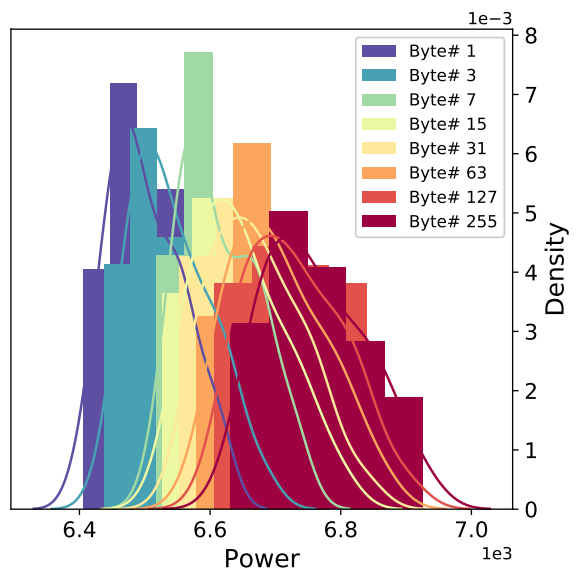

(d) Power consumption distribution for arbitrarily chosen data stored in a cache set.

Fig. 6: Leakage in Shakti-C data cache due to cache sets and data. The former being the more prominent of the leakages.

**Leakage in Register Files and Interstage Buffers.** For a load instruction, data in MEM-WB buffer is forwarded to Register File (RF) (or vice-versa for a store instruction) as shown in Figure 4. The Physical Register File (PRF) stores the output from the execution unit during operand forwarding and also during a cache miss.

The pipeline buffer ID-EXE lies in between decode and execute stage. It holds the data read from the RF or PRF. Similarly, the EXE-MEM and MEM-WB buffers, stores the results of the execute stage and the data corresponding to the load and store instruction. All these modules directly hold sensitive data, therefore, have high SVF. Additionally, the Control and Status Register (CSR) has a high SVF of 0.64.

**Leakage in ALU.** For an ALU operation, register operands from the RF are fetched and stored in the ALU register. For instance, operations such as $p \oplus k$ (Line 5 in Figure 5), simple addition having sensitive information as in Line 7 results in high SVF in the ALU modules.

**Leakage due to EDA Tool Translations.** A hardware design flows through various EDA tools during the design process. Each tool performs several translations on the design. For example, tools may optimize the design to reduce area, energy, or to improve performance. Each of these translations is done ensuring that the functionality of the design is intact. However, none of the EDA tools evaluate for side-channel security. As a result, the translations may increase the side-channel vulnerability of the device.

The Shakti-C [12] RISC V processor is coded in Bluespec

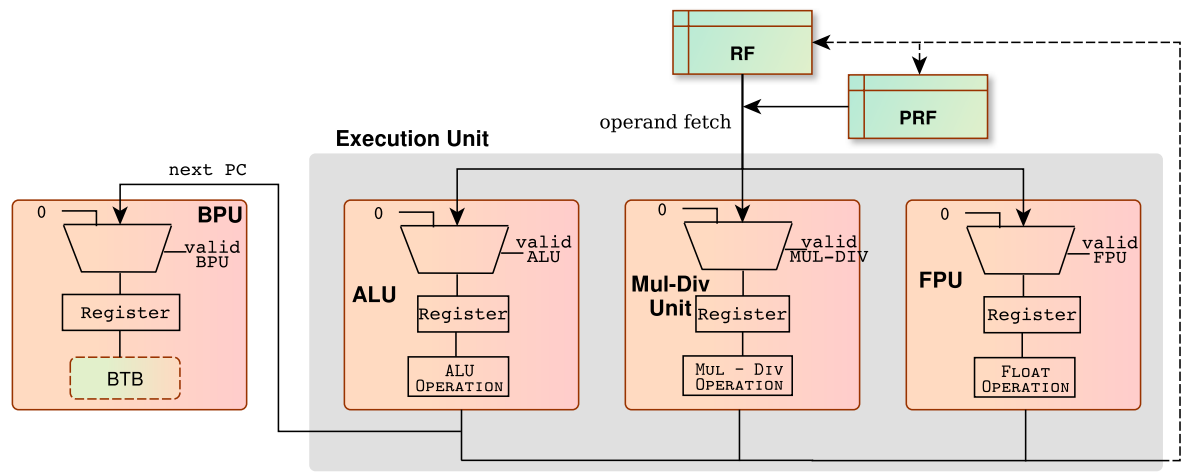

(a) Expected design of Execution Unit in Shakti-C.

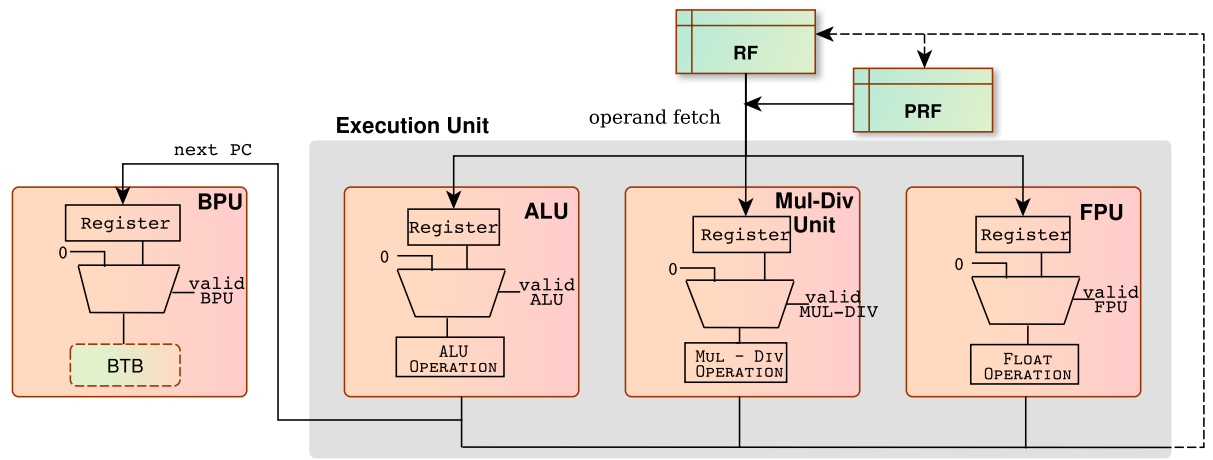

(b) Design after Bluespec compilation. The change in the placement of register causes increased leakage.

Fig. 7: EDA tools perform translations keeping functionality intact. Some of these translations may increase side-channel vulnerability as we found in the Execute Unit of Shakti-C when compiled with Bluespec.

System Verilog (BSV) [16]. The BSV code is compiled by the Bluespec.2018.10.beta1 compiler generating synthesizable Verilog RTL. We look at the Execute Unit of Shakti-C to understand how the Bluespec compiler increases side-channel vulnerability.

Figure 7a shows the expected behavior of the Execution Unit of Shakti-C. In the execute stage, register operands are read from the Register File (RF) or PRF and forwarded to the different functional units, such as the ALU, FPU, and Mul-Div. In each of the units, a multiplexer chooses to forward the operands to a register depending on the select line. For instance, for an ALU operation, register operands are only sent to the register in the ALU. The registers in the other functional units will store the default value 0.

Figure 7b shows the Execute Unit of Shakti-C after the compilation of the Bluespec code. The positions of the multiplexer and register in each functional unit are interchanged. While this modification does not change the functionality of the design, we observe that the register operands are stored in all units independent of the type of operation. Therefore, register operands corresponding to an ALU operation are stored in FPU and Mul-Div modules. Thus, for the AES-128 benchmark, in spite of having no floating point operations, multiplications or divisions, PLAN shows a high SVF in the FPU and Mul-Div units as seen in Table I and Figure 3.

A similar observation is also made in the branch prediction unit (BPU). The result of the ALU is forwarded to the BPU to store the target of a branch instruction. As with the FPU and Mul-Div units, the position of the Multiplexer and BPU register are interchanged by the Bluespec compiler. Thus, every output of the ALU gets stored in the BPU register, causing it to leak. Hence the BPU shows up with high SVF (see Table I and Figure 3) for the AES-128 benchmark even though there are no sensitive branch operations. These leakages can be fixed by minor RTL changes that prevent the compiler from placing the register ahead of the multiplexer. Section IV provides more details.

### B. Caveats on PLAN

PLAN is designed to provide a fine-grained evaluation of the power side-channel leakage in hardware designs. It is designed to provide quick results and therefore its scope is restricted to the most important leakages. In spite of the reduced scope, we show that our hardened processor, PARAM, withstands Differential Power Analysis (DPA), even after one million power traces. In this section, we enumerate the caveats of PLAN.

**Linear Correlation.** SVF measures the linear correlation between Oracle and Side-Channel vectors. Most powerful side-channel attacks such as DPA, exploit such linear correlations. SVF cannot identify leakages when the Oracle and Side-Channel vectors are non-linearly correlated.

**Leakage through Static Power Consumption.** PLAN is restricted to evaluating leakage due to the dynamic power consumption of the processor. The dynamic power consumption is the most exploited in power side-channel attacks. Recently, information has shown to leak from the static power consumption [17], [18]. These attacks are considerably more difficult as the leakage is several times smaller than dynamic power consumption. Leakage due to static power consumption is beyond the scope of this work.

**Use of Pre-synthesized Netlist.** Pinpointing leaking modules and characterizing the leakage cannot be done on actual hardware. It requires a simulated environment and access to the device's source code.

PLAN evaluates leakage based on behavioral simulation. Power consumption variations due to placement and routing, and other timing constraints are therefore not considered. The advantage we gain by using behavioural simulation is that the time required for executing PLAN is significantly lesser compared to when PLAN uses a post place-and-route simulation. For example, evaluating the complete Shakti-C processor using behavioral simulation takes around 5 hours on a standard Intel i5 desktop CPU. Evaluating just the data cache of Shakti-C with post place-and-route simulation, takes over 9 hours in the same environment. A single evaluation of the complete Shakti-C is expected to take over a month.

**Detecting Leakages due to EDA Translations.** Every EDA translation can introduce new side-channel vulnerabilities. Since PLAN works with the pre-synthesized netlist, leakage detection due to the EDA tool is restricted only to this stage. Vulnerabilities that may be introduced in other EDA translations are not identifiable.

## IV. The Design of a Power Attack Resistant Microprocessor: PARAM

To design a power side-channel attack resistant microprocessor would require all leaking modules to be fixed. From the observations in Section III-A, leaking modules can be categorized as (1) modules that hold data correlated with the secret and (2) modules that leak due to security agnostic EDA translations. In this section, we harden Shakti-C by either reducing or eliminating leakages. The resultant, power attack hardened processor is called PARAM.

### A. Preventing leakage due to data correlated with the secret

Modules in the data path such as Register Files, Inter Stage buffers, Control Status words, and the memory hierarchy store data that is highly correlated with the secret. Hence these modules contribute considerably to the side-channel leakage of the processor. Further, in the cache memory, we find that the fixed mapping from address to cache set causes significant leakage. In this section, we introduce obfuscation techniques to reduce these leakages.

**Reducing leakage in the data path.** To reduce leakage in modules present in the data path, we design PARAM such that data is almost always in an obfuscated form. The obfuscation breaks the correlation between the secret and power consumed by the module thus reducing leakage. When data, $d$, is fetched from the off-chip memory due to a cache miss, it is obfuscated using a tiny algorithm, $O_k$, with secret key $k$. The secret key is generated securely within the processor and never exposed. The obfuscation operation can be represented as follows:

$$d' = O_k(d) \ . \tag{5}$$

The obfuscated data ($d'$) is stored in all storage elements in the data path, including the data cache, LB and HB, Pipeline Buffers, Memory Access Units, and Register files (RF and PRF). In Figure 8, this obfuscated data path is shown in red dashed lines. The obfuscated data $d'$ is de-obfuscated using the inverse function $O_k^{-1}$ as follow:

$$d = O_k^{-1}(d') \ . \tag{6}$$

De-obfuscation is done when an operation needs to be performed on the data, typically in the execute stage of the processor. The results of the operation are encrypted again before it is passed through the rest of the pipeline. De-obfuscation also occurs during a cache eviction, when data is written back to the off-chip RAM.

**Reducing Leakage due to fixed address to cache set mapping.** As seen in Figure 6, each cache set has a different power consumption signature. To hide this signature, PARAM obfuscates the address, $a$, sent to the cache memory. Obfuscation is done for the tag and set index bits, keeping the offset bits unchanged. Let $\texttt{tagsetindex}(a)$ denote the tag and set index bits of an address $a$ and $\texttt{offset}(a)$ denote the offset bits of the address $a$, then the obfuscation of address $a$, denoted $a'$, is done as follows:

$$a' = O_k(\texttt{tagsetindex}(a))||\texttt{offset}(a) \ , \tag{7}$$

where $||$ is the concatenation operation. Thus, for every request to the data cache, the tag and set index bits of the address are first encrypted. This obfuscation breaks the deterministic address to cache set mapping. Further, the key $k$ is changed periodically. For every key, the mapping from address to cache set changes. Thus, the same address gets mapped to different cache sets over time. This prevents the attacker from learning the unique power consumption signatures of a cache set. Figure 9 shows the t-scores computed on pairwise power consumption between cache sets. Compared to Figure 6, the differential power consumption (obtained by Hamming distance model) is reduced considerably indicating significantly less leakage. Later in the section, we provide more details about this remapping.

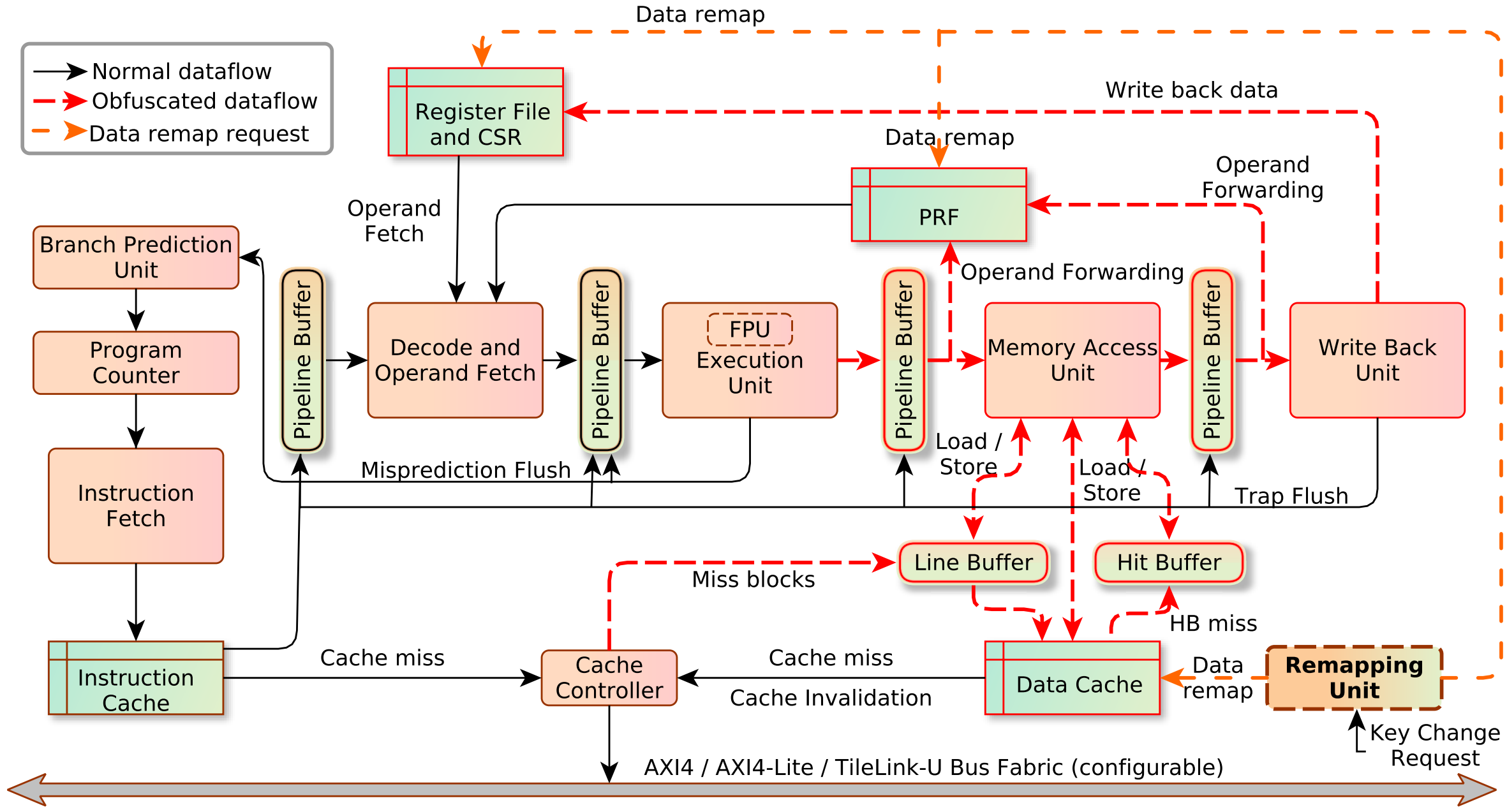

Fig. 8: Overview of PARAM processor architecture having protection against power side-channel attacks by obfuscating address lines of data cache and data of data cache memory subsystem, Physical Register File (PRF) and Register File. Remapping unit is responsible for updating data values in the above mentioned locations whenever the obfuscation key changes. Red dashed lines are carrying obfuscated data values from which the actual data can be retrieved by using key values.

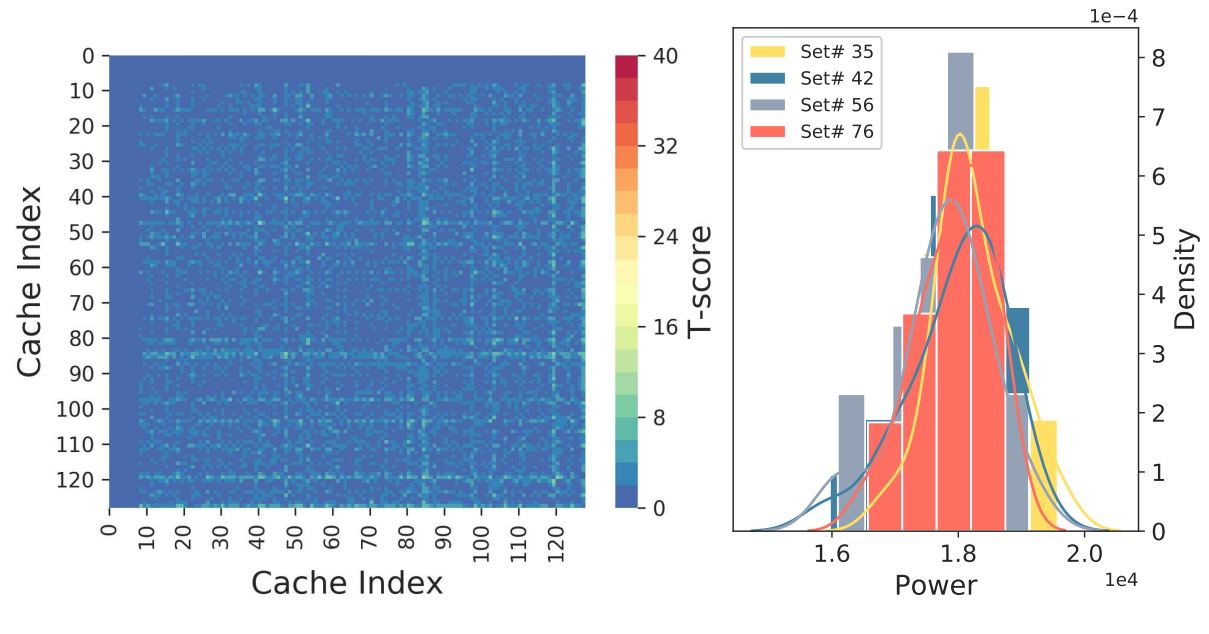

(a) (b)

Fig. 9: Address and data obfuscation in the data cache reduces leakage considerably. The power consumption distributions of arbitrary cache sets are almost similar.

**Obfuscation Function.** The main purpose of the obfuscation function is to break the correlation between the power consumption of the processor and the data stored. The obfuscation function is in the critical path, therefore this function has to be extremely small. We choose a 4-round Feistel structure to perform the obfuscation. The input address or data to be obfuscated is of 32 bits and is split into two parts of 16 bits each. The Feistel structure takes 16 key bits, generated from an LFSR present in the hardware as input in every round and uses an Affine transformation to combine the key with the input. The output is a well-mixed function of the input and key. Appendix A has more details on this obfuscation function.

*Security Evaluation.* All keys are generated internally and completely transparent from the users. Similarly, all obfuscated data is not accessible by any user of the processor. Due to this, the only possible attack is through side-channel leakage of the key. To protect against side-channel attacks, we adopt a re-keying countermeasure [19] by which the key is changed at regular intervals to ensure that an attacker does not obtain sufficient side-channel information to break the obfuscation. The re-keying requires a remapping of the obfuscated data and addresses to ensure correct functionality. We next discuss the Remapping Unit present in PARAM.

**Remapping Unit.** The Remapping Unit is centralized in PARAM (see Figure 8). Upon receiving a key change request from the software, it first generates a new key from an LFSR present in the hardware. The new key is used to remap the obfuscated data. This is done by de-obfuscating with the old key ($k_o$) and then obfuscating again with the new key ($k_n$) as follows:

$$d'' = O_{k_n}(O_{k_o}^{-1}(d')) \ . \quad (8)$$

Additionally, since the mapping from address to cache sets are obfuscated, PARAM remaps the cache sets based on the new key. A straightforward way to achieve this is to invalidate the entire cache. This involves, writing back dirty cache lines to the off-chip RAM and then changing the obfuscation key. However, this requires the data cache to be blocked during the remapping phase. Further, there would be an increase in the number of cache misses, since the data has to be reloaded into

(a)

```
rule rl_addition;
  if(a1 > 'h20)
    rg_out <= tagged Valid (a1+a2);
  else
    rg_out <= tagged Invalid;
Endrule
/* A combinational block then reads
rg_out and checks valid bit before
using it */
```

BSV to Verilog Translation by Bluespec Compiler

```
wire rg_out$D_IN;
reg rg_out;

assign rg_out$D_IN =
{a1>'h20, a1+a2};

always@(posedge clk) begin
  rg_out <= rg_out$D_IN;
end

assign out = rg_out[31:0];
assign RDY_out = rg_out[32];
```

(b)

```
rule rl_addition;
  if(a1 > 'h20)
    rg_out <= {1'b1, (a1+a2)};
  else
    rg_out <= {1'b0, 32'd-1};
Endrule
/* A combinational block then reads
rg_out and checks valid bit before
using it */
```

BSV to Verilog Translation by Bluespec Compiler

```
wire rg_out$D_IN;
reg rg_out;

assign rg_out$D_IN = a1 > 'h20 ?
{1'b1, a1+a2} : 33'h0FFFFFFFF;

always@(posedge clk) begin
  rg_out <= rg_out$D_IN;
end

assign out = rg_out[31:0];
assign RDY_out = rg_out[32];
```

Fig. 10: An illustration of leakages due to EDA translations (a) Before Fix (b) After Fix

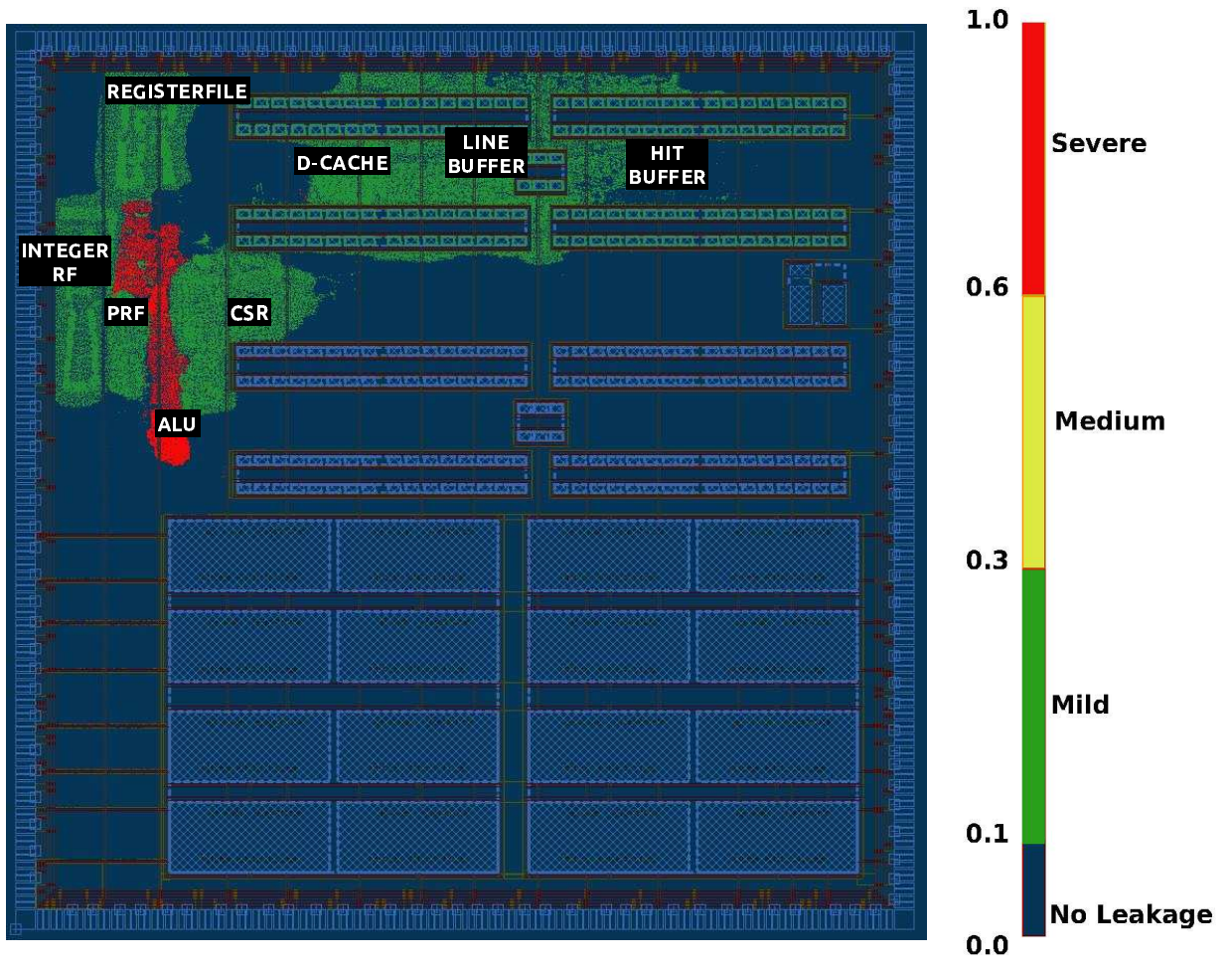

Fig. 11: Information leakage plot shows the floor plan for PARAM illustrating the modules with their SVF values in different colors.

the cache. An alternative was suggested by Qureshi in [20], where a gradual remap was used to reduce overheads. In the gradual remap, a cache set will be remapped periodically without blocking the cache. PARAM uses the former approach. Qureshi's gradual remapping will be incorporated in the future.

### *B. Preventing power leakages due to EDA translations.*

Shakti-C is coded using Bluespec System Verilog (BSV). We explain the leakage caused by the Bluespec translation from BSV to synthesizable Verilog, with a small example of an addition. In Figure 10a, ideally, the register should have been latched with the output of the adder (a1 + a2) only when the condition (a1 > 'h20) is true. However, to save a multiplexer and the associated delay, we see that the EDA translation removes this check and appends the $33^{rd}$ bit with the condition (a1 > 'h20). Now, the adder's outputs are always latched to the register but discarded when it is read by some other combinational logic, that first checks the $33^{rd}$ bit for validity. Although this is functionally correct, it leaks information as the result of the adder's outputs are always latched.

To stymie this leakage, we modified the Bluespec code, added the valid bit and made sure that a constant value 1 is latched onto the register when the condition fails, as illustrated in Figure 10b. This small change, that costs an extra multiplexer, reduces leakage. In the context of Shakti-C, the datapath from the register-file to FPU, Mul-Div unit and the datapath from ALU to BPU have been altered to send data only when the data is associated with the corresponding units, similar to the example illustrated here.

## V. Experimental Validation

To validate the information leakage in PARAM, we passed PARAM through PLAN. We found that leakages due to correlation between sensitive data and power consumption reduced in memory hierarchy modules and register file. Figure 11 corroborates our result. Information leakage due to EDA tool translations in modules such as BPU, FPU and Mul-Div unit have been completely eliminated in PARAM. However, leakage due to the ALU module is not altered. We synthesized PARAM for a Xilinx Kintex-7 FPGA (xc7k160tfbg676-1). Compared to Shakti-C, which had (51K LUTs and 21K FFs), PARAM requires 73K LUTs and 22K FFs, an increase of 22K more LUTs. The maximum clock frequency reduced from 48MHz in Shakti-C to 33MHz in PARAM. The increase in area and clock period is mainly due to the obfuscation circuit. Each realization of the obfuscation function is a combinational logic requiring 152 LUTs. Several instances of this circuit are needed. More efficient obfuscation functions can reduce these overheads considerably. It should be noted, that since all additions to PARAM were combinational circuits, the cycles per instruction (CPI) for the processor is the same as that of Shakti-C.

To test the efficacy of PARAM at withstanding side-channel attacks, we mounted a first-order DPA with AES-128 executing on the processor and compared the results against the original Shakti-C processor. The DPA test platform comprised of a SASEBO-GIII FPGA board which instantiated the processors and Lecroy Teledyne HDO4104 oscilloscope. The DPA results are shown in Figure 12. The correct key value is disclosed after 62319 traces in the unprotected Shakti-C processor. On the other hand, in PARAM no AES key was recovered even after 1 million power traces (see Figure 12b). Figure 13 shows the correlation scores for Shakti-C and PARAM. It can be seen that in Shakti-C, the correct key is highly correlated compared to the wrong keys. However, no such high correlation is observed in PARAM.

## VI. Related Work

Since Paul Kocher's seminal work on Differential Power Analysis (DPA) [2], there have been innumerable attempts to protect devices against the attack. While algorithmic level countermeasures such as [19], [21], customize protocols and cipher algorithms to be inherently secure, implementation level countermeasures, such as masking [7], [22] and threshold implementations [23], use randomization to hide leakage due to sensitive operations. All these approaches are designed to protect crypto-algorithms and cannot be generalized for an arbitrary application. Device-level countermeasures like [8] use custom gates that consume power independent of the gate's switching. While these approaches can be generalized to protect any application, their performance overheads, typically over 100%, limits their usability to very small circuits. This approach is therefore not suitable to protect entire microprocessors. Recently, Nabel et al. [24] proposed a processor that operates on encrypted data and supports arithmetic operations such as modular multiplication, exponentiation, and inversion. In addition to other security benefits, their processor is resistant to side-channel attacks. However, the processor has limited capabilities, such as in Root-of-Trust devices. PARAM, on the other hand, is the first work to achieve side-channel security in general purpose microprocessor. Besides providing blanket security to all applications running on the processor, the overheads are lower compared to other countermeasures.

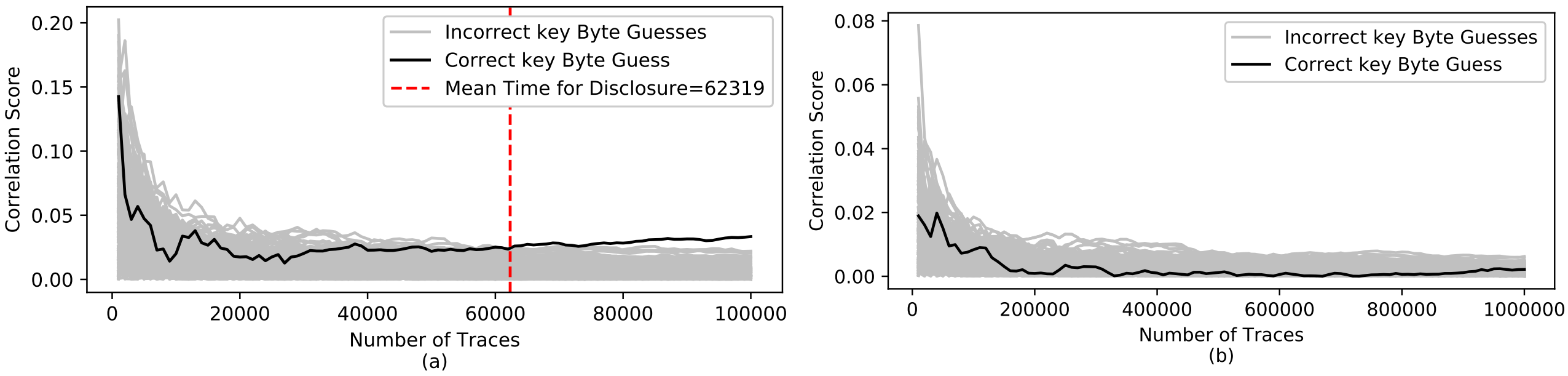

Fig. 12: DPA results of reference architecture before and after leakage mitigation. a) Shakti-C architecture has Mean Time for key Disclosure at 62319 traces b) PARAM has very low correlation score for correct key byte guess even for 1 million traces.

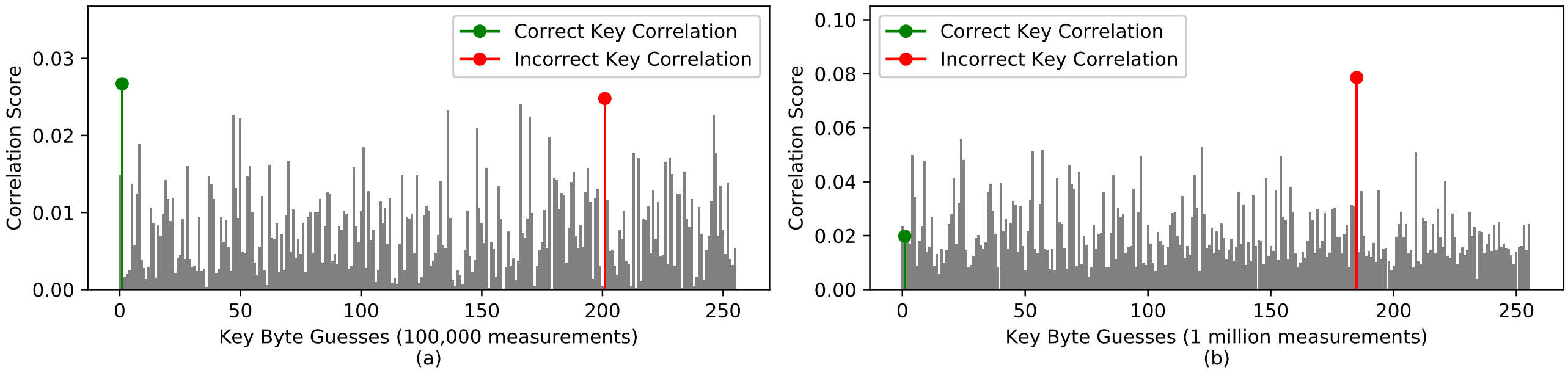

Fig. 13: Correlation scores for each key byte guess in DPA. a) Shakti-C has high correlation for correct key byte value for 62319 traces. b) Correlation scores of key byte guess in PARAM architecture for 1 million traces.

## VII. Limitations of PARAM

While we emphasize the fact that PARAM is the first general purpose processor hardened for power side-channel attacks, it has some limitations, which we will now enumerate.

- PARAM is designed by fixing the leaking modules that PLAN identifies. All the limitations of PLAN (discussed in Section III-B) are inherent limitations of PARAM. Thus, for example, PARAM cannot protect against attacks that use leakage through static power consumption.
- Most leakages in Shakti-C have either been eliminated or reduced considerably. However, some modules still leak, for example in the ALU. Protecting the ALU requires all arithmetic operations to be protected. This is part of our future work.
- Our focus in this paper was to reduce leakage in the data path of the processor. Leakage can also occur in the control path. For example, consider the code snippet shown below which computes $\mathtt{a}^{\mathtt{key}}$.

```
exp(a, key){
   for(r=1, i=0; i < 1024; ++i){
      if (key[i] == 1) r = r * a;
        r = r * r;
    }
    return r;
}
```

Leakage about the key occurs due to the differential paths taken when $\mathtt{key} = 1$ and $\mathtt{key} \neq 1$. We believe that these control path leakages can be better solved with the help of compiler techniques. Obfuscation in the hardware alone may not be able to protect against these leakages.

## VIII. Conclusion

This paper proposes the first general purpose microprocessor with built-in security for power side-channel attacks. The processor, which we call PARAM, is derived from a RISC V processor, with a majority of the side-channel vulnerabilities rectified. During the process of this design, we found interesting causes of side-channel leakage such as the EDA tool translations and address to cache set mapping. A combination of programming techniques and obfuscation of the data path helped reduce the leakage considerably. PARAM is validated for correctness on an FPGA and its side-channel security verified by mounting a first-order DPA attack. While PARAM's overheads are low compared to other DPA countermeasures, we believe that the overheads can be reduced further by better designs for the obfuscation functions and by increasing the number of pipeline stages in the processor. This is left as future work.

## Appendix A
## Obfuscation Function

PARAM uses a 4-round Feistel network having an Affine function $F$ in each round to transform the data and key values as shown in Figure 14. In each round, the input is divided into two halves, say $L$ and $R$, in which the $L$ part is ex-ored with the output of the Affine function $Y$. The Affine function takes the $R$ part and the key as inputs, each of $n$ bits, and produces an $n$ bit output as follows:

$$Y = A(R||K) + C \ , \tag{9}$$

where $A$ is an affine matrix and $C$ is a round constant. The Affine function for the first round with $n = 16$ is shown in the Figure 15. In the next round, R and output of ex-or are swapped and fed as L and R respectively. To perform de-obfuscation keys are fed in the reverse order.

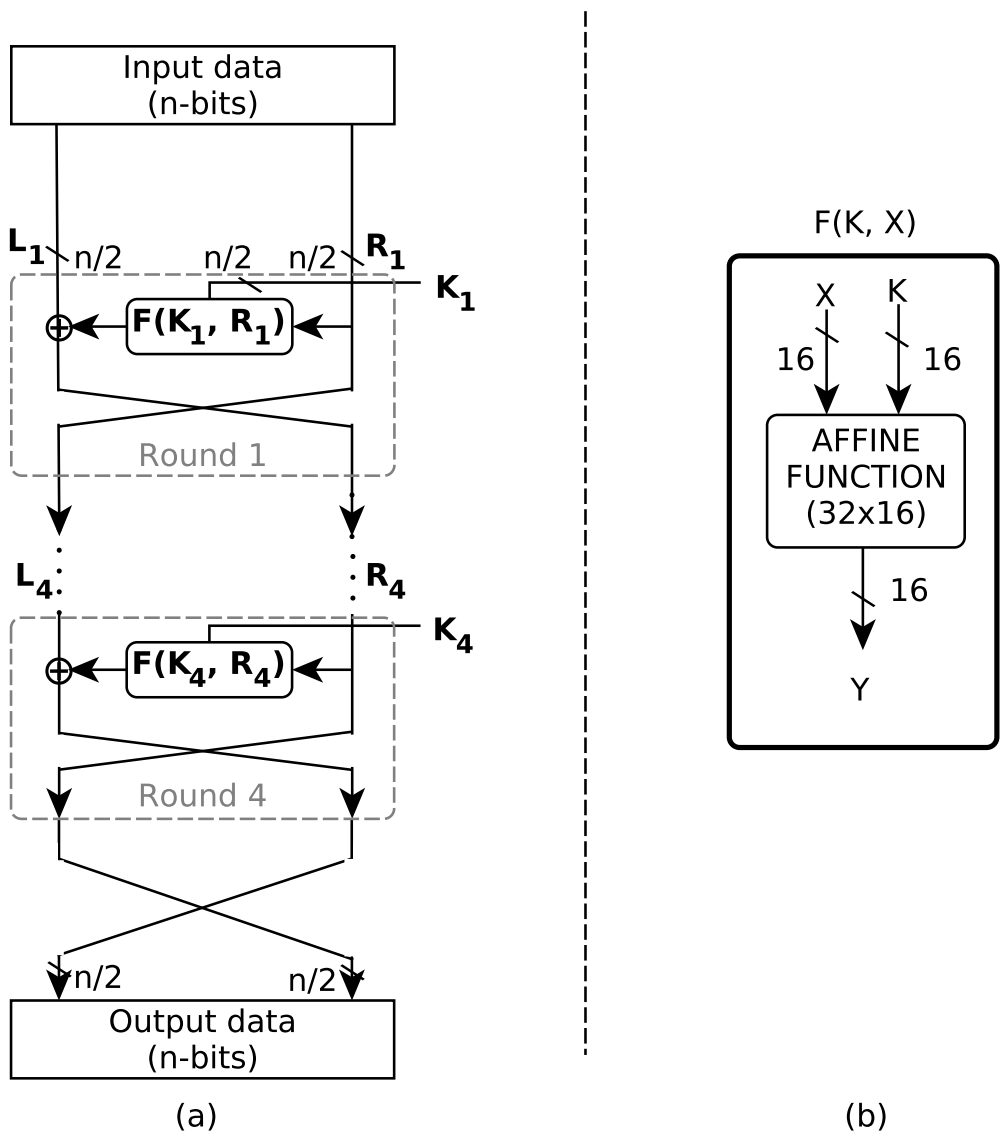

Fig. 14: (a) Structure of the obfuscation function uses Feistel network with 4 rounds. (b) 32×16 affine function used in Feistel structure.

$$
\begin{bmatrix} y_0 \\ y_1 \\ y_2 \\ y_3 \\ y_4 \\ y_5 \\ x_6 \\ y_7 \\ y_8 \\ y_9 \\ y_{10} \\ y_{11} \\ y_{12} \\ y_{13} \\ y_{14} \\ y_{15} \end{bmatrix}
=
\begin{bmatrix}
1&0&0&0&0&0&1&1&1&1&0&1&1&1&1&1&1&0&0&0&1&1&1&0&0&0&1&1&0&1&1&0\\
0&1&0&0&1&1&1&0&1&1&0&0&0&0&1&0&0&0&0&1&0&1&1&1&0&0&1&0&1&0&1&0\\
0&1&1&0&0&1&1&1&0&1&0&0&0&0&0&0&0&0&1&1&1&0&0&0&0&0&0&0&1&0&0&1\\
0&1&1&1&0&0&1&0&0&1&1&0&0&1&1&1&1&1&0&1&1&1&0&1&1&0&1&1&1&0&1&1\\
1&1&1&1&1&0&1&0&1&1&1&0&0&0&1&0&1&1&1&1&0&0&0&0&1&0&1&1&0&1&1&0\\
0&0&0&1&0&0&1&0&1&1&1&0&0&1&0&1&0&1&1&0&0&1&0&1&1&1&1&1&1&1&1&1\\
0&1&0&1&0&0&0&1&0&0&1&0&0&1&0&0&1&0&0&0&0&1&0&1&0&0&0&1&1&0&1&1\\
0&1&0&1&0&0&1&0&0&0&0&1&0&1&0&0&0&0&0&1&1&0&0&1&1&1&1&1&0&0&0&0\\
0&0&0&1&0&0&0&0&0&1&0&1&0&1&0&1&1&0&0&0&0&1&1&1&0&0&0&1&1&0&1&1\\
0&1&1&1&1&1&0&0&1&1&0&1&1&1&1&1&0&0&0&1&1&1&1&0&0&0&1&1&1&0&0&0\\
1&1&1&0&0&0&1&1&1&0&1&0&0&1&1&0&1&1&1&1&0&0&1&1&0&1&0&0&1&1&0&0\\
0&0&1&1&1&0&1&0&1&0&1&1&1&1&0&0&0&0&0&1&1&0&1&1&0&1&1&0&1&1&1&1\\
1&0&0&0&1&1&0&1&0&0&0&1&1&1&0&1&1&0&1&1&0&0&1&0&1&0&0&0&1&0&1&0\\
1&0&0&1&0&1&0&0&0&0&1&1&0&1&1&0&0&1&0&1&1&1&1&1&0&0&1&1&1&1&0&1\\
1&1&1&0&1&1&1&0&0&0&0&1&0&0&1&0&0&1&1&0&1&1&1&1&1&1&1&1&0&1&0&1\\
0&1&1&0&1&1&1&0&0&1&0&1&0&0&1&1&0&1&1&1&0&0&1&1&1&1&0&1&0&1&1&0
\end{bmatrix}
\cdot
\begin{bmatrix} x_0 \\ x_1 \\ x_2 \\ x_3 \\ x_4 \\ x_5 \\ x_6 \\ x_7 \\ x_8 \\ x_9 \\ x_{10} \\ x_{11} \\ x_{12} \\ x_{13} \\ x_{14} \\ x_{15} \\ x_{16} \\ x_{17} \\ x_{18} \\ x_{19} \\ x_{20} \\ x_{21} \\ x_{22} \\ x_{23} \\ x_{24} \\ x_{25} \\ x_{26} \\ x_{27} \\ x_{28} \\ x_{29} \\ x_{30} \\ x_{31} \end{bmatrix}
+
\begin{bmatrix} c_0 \\ c_1 \\ c_2 \\ c_3 \\ c_4 \\ c_5 \\ c_6 \\ c_7 \\ c_8 \\ c_9 \\ c_{10} \\ c_{11} \\ c_{12} \\ c_{13} \\ c_{14} \\ c_{15} \end{bmatrix}
$$

Fig. 15: Affine transformation used in the obfuscation for the First Round.